\documentclass[prd,onecolumn,aps,superscriptaddress,nofootinbib,hidepacs]{revtex4}
\usepackage[]{epsfig}
\usepackage[centertags]{amsmath}
\usepackage{amsfonts}
\usepackage{amssymb}
\usepackage[english]{babel}
\newcommand{\be}{\begin{equation}}
\newcommand{\ee}{\end{equation}}
\newcommand{\ba}{\begin{eqnarray}}
\newcommand{\ea}{\end{eqnarray}}

\begin{document}

\title{String solutions in Chern-Simons-Higgs model coupled to an axion}

%%%%%%%%%%%%%%%%%%%%%%%%%%%%%%%%%%%%%%%%%%%%%%%%%%%%%%%%%%%%%%%%%%%%%%%%%%%%%%%
\author{ J.~L\'opez-Sarri\'on}
\affiliation{Departamento de F\'{\i}sica,  Universidad de Santiago
de Chile\\
Casilla 307, Santiago, Chile}
%%%%%%%%%%%%%%%%%%%%%%%%%%%%%%%%%%%%%%%%%%%%%%%%%%%%%%%%%%%%%%%%%%%%%%%%%%%%%%%
\author{E.~F.~Moreno}
\thanks{Associated with CONICET}
\affiliation{Departamento de F\'{\i}sica, Facultad de Ciencias
Exactas\\
Universidad Nacional de La Plata, C.C 67, 1900 La Plata,
Argentina}
\affiliation{Department of Physics, West Virginia
University\\
Morgantown, West Virginia 26506-6315, U.S.A.}
%%%%%%%%%%%%%%%%%%%%%%%%%%%%%%%%%%%%%%%%%%%%%%%%%%%%%%%%%%%%%%%%%%%%%%%%%%%%%%%
\author{F.~A.~Schaposnik}
\thanks{Associated with CICBA}
\affiliation{Departamento de F\'{\i}sica, Facultad de Ciencias
Exactas\\
Universidad Nacional de La Plata, C.C 67, 1900 La Plata,
Argentina}
%
%%%%%%%%%%%%%%%%%%%%%%%%%%%%%%%%%%%%%%%%%%%%%%%%%%%%%%%%%%%%%%%%%%%%%%%%%%%%%%%
\author{D.~Slobinsky}
\affiliation{Departamento de F\'{\i}sica, Facultad de Ciencias
Exactas\\
Universidad Nacional de La Plata, C.C 67, 1900 La Plata,
Argentina}
%%%%%%%%%%%%%%%%%%%%%%%%%%%%%%%%%%%%%%%%%%%%%%%%%%%%%%%%%%%%%%%%%%%%%%%%%%%%%%%

%%%%%%%%%%%%%%%%%%%%%%%%%%%%%%%%%%%%%%%%%%%%%%%%%%%%%%%%%%%%%%%%%%%
\begin{abstract}
We  study a $d=2+1$ dimensional Chern-Simons gauge theory
coupled to a Higgs scalar and an axion field, finding the form of
the potential that allows the existence of selfdual equations and
the corresponding Bogomolny bound for the energy of static
configurations. We show that the same conditions allow for the
$N=2$ supersymmetric extension of the model, reobtaining the BPS
equations from the supersymmetry requirement. Explicit
electrically charged vortex-like solutions to these equations are
presented.
\end{abstract}


 \maketitle
%%%%%%%%%%%%%%%%%%%%%%%%%%%%%%%%%%%%%%%%%%%%%%%%%%%%%%%%%%%%%%%%%%%

\section{Introduction}
There has been recently a revival in the interest on
Abrikosov-Nielsen-Olesen flux tubes in connection with
supergravity models and the realization of cosmic superstrings.
Indeed, although  such cosmic size superstrings were originally
excluded in the context of perturbative string theory \cite{witt},
it became clear recently that the question  should be revisited as
the relevance of branes and new kind of extended objects was
understood (see \cite{pol} and references therein). In order to
understand the nature and structure of such {\it stringy} cosmic
strings, the embedding  of BPS objects in supersymmetry and
supergravity models has become an active area of research so that
the properties of BPS solitons, their connections with the
supersymmetry algebra and their cosmological applications have
been discussed by many authors \cite{ens1}-\cite{Gorsky}.

Having in mind the study of BPS solitons in a string theory
context,  it is natural to consider models where an axion field is
included. In particular, an  $N=1$ globally supersymmetric model
in $d=3+1$ dimensions consisting of an axion superfield $S$
coupled to $W_\alpha W^\alpha$, with $W_\alpha$ the chiral
superfield strength, was analyzed in  \cite{Blanco} and finite
energy cosmic string solutions were constructed. Also, the impact
of axions on dynamics of a $d=3+1$ Yang-Mills theory supporting
non-Abelian strings has been analyzed in \cite{Gorsky}.

It is the purpose of this work to consider similar issues in
 $d= 2+1$ space-time dimensions for which a rich variety of flux
tube solutions exists already when the axion field is absent.
Indeed, when gauge fields with dynamics governed by a Chern-Simons
action are coupled to charged scalars with an appropriate sixth
order symmetry breaking potential, the model admits BPS equations
with vortex-like solutions carrying both magnetic flux and
electric charge \cite{Hong}-\cite{JW}. It should be stressed that
in the absence of the Chern-Simons term, electrically charged
vortices with finite energy (per unit length) do not exist
\cite{JZ}. Hence, the model we are interested in, could show novel
aspects of charged string like configurations when an axion is
present, in particular with respect to their application to
cosmological problems. This in the perspective that at high
temperatures, a relativistic four dimensional quantum field theory
becomes effectively three dimensional.

The paper is organized as follows:  we start by considering, in
section 2 a purely bosonic model with couplings and potentials
chosen so as to have a gauge invariant action leading to
non-trivial Bogomolny equations. These equations are the natural
extension,  when an axion field is present, of those found in
\cite{Hong}-\cite{JW}. We present in section 3 the supersymmetric
extension of the model and we establish the connection between
supersymmetry and BPS equations when the axion field is included.
Numerical solution of such first order coupled equations are
presented in section 4, where we discuss, in particular, novelties
in the vortex solutions resulting from the presence of the axion
field. A summary and discussion of our results are presented in
section 5.

\section{Axion coupled to a Chern-Sim\-ons-Higgs system}

The coupling of an axion to a gauge field with dynamics governed
by a Chern-Simons action poses some problems
\cite{Burgess}-\cite{FG}. To discuss how can they be overcome, let
us consider the following $(2+1)$-dimensional bosonic action,

\begin{equation}
{\cal S} = \int d^3x\, \left\{\frac{\kappa}{8\delta}f(s){\rm
Im}D_\mu S\tilde F^\mu + \vert D_\mu \phi\vert^2 +
K^{\prime\prime}(s) \vert D_\mu S\vert^2 - W(\phi, S)\right\}
\label{accion}
\end{equation}
where $\phi$ and $S = s + ia$ are complex fields, $A_\mu$ is a
$U(1)$ gauge field, and $\tilde F^\mu$ is defined as,
\begin{equation}
\tilde F^\mu \equiv \epsilon^{\mu\nu\sigma}\partial_\nu A_\sigma
\end{equation}
$W(\phi,\phi^*,S,S^*)$ is a potential term, $f$ and $K$ are
arbitrary functions of $s$, the real part of $S$, primes stand for
$\partial/\partial s$, $\kappa$ is a constant and $\delta$ is a
dimensionless parameter (which in the 4 dimensional case is
related to the Fayet-Iliopoulos term and the Planck mass).
Finally, $D_\mu$ is the covariant derivative acting on the fields
$\phi$ and $S$ according to
\begin{eqnarray}
D_\mu \phi &=& \partial_\mu \phi - ieA_\mu\phi\nonumber\\
D_\mu S &=&  \partial_\mu S + 2i\delta A_\mu
\end{eqnarray}
As done in \cite{Blanco} for the $3+1$ model, we shall identify
$a$ in (\ref{accion}) with the axion field and $s$ with a dilaton
field. Note that because of the definition of the axion covariant
derivative,  the Chern-Simons term appears in action
(\ref{accion}) multiplied by the factor $f(s)$,
\begin{eqnarray}
{\cal S}_{CS}[A,s] = \frac{\kappa}{4} \int d^3x f(s)
 \epsilon^{\mu\nu\sigma} A_\mu \partial_\nu A_\sigma
 \label{normal}
\end{eqnarray}

The action (\ref{accion}) is invariant under the local
transformation
\begin{align}
S ~\; &\to \; S - 2i\delta \Lambda(x)\nonumber\\
S^* \; &\to \; S^* + 2i\delta \Lambda(x)\nonumber\\
\phi ~\; &\to \; e^{ie\Lambda(x)}\phi\label{gauge}\\
\phi^* \; &\to \; \phi^* e^{-ie\Lambda(x)}\nonumber\\
A_\mu \; &\to \; A_\mu + \partial_\mu \Lambda(x)\nonumber
\end{align}

The time component of the gauge field equation of motion is the
Chern-Simons version of the Gauss law and can be used to solve for
$A_0$ giving
\begin{equation}
A_0 = \frac{\kappa\,{\cal B}}{2(e^2\vert\phi\vert^2 +
4\delta^2K^{\prime\prime})} \label{gausi}
\end{equation}
where
\begin{equation}
{\cal B} \equiv f(s)F_{xy} - \epsilon^{ij}(A_i +
\frac{1}{2\delta}\partial_ia)\partial_jf(s)
\end{equation}
The energy can be found from the energy-momentum tensor obtained
by varying the action with respect to the metric,
\begin{equation}
\delta S = \frac{1}{2} \int d^3x \sqrt g \, T^{\mu\nu} \delta
g_{\mu\nu}
\end{equation}
Integration of the time-time component $T^{00}$ gives
\begin{equation}
{\cal E}= \int d^2x\,\left\{ \vert D_i\phi\vert^2 +
K^{\prime\prime}\vert D_i S\vert^2 + W(\phi,S) + \frac{\kappa^2
{\cal B}^2}{4(e^2\vert\phi\vert^2 +
4\delta^2K^{\prime\prime})}\right\}
\end{equation}
After some work, this expression can be written in the form
\begin{align}
& \hspace{-1 cm} {\cal E} = \int \!\!d^2x\!\!\left(
\!\vphantom{\frac{\left(e^2\vert\phi\vert^2 + 4\delta^2
K^{''}\right)}{s/2}} \vert D_x\phi\pm iD_y\phi\vert^2 +
K^{\prime\prime}\vert D_xS \pm iD_y S\vert^2 +
\frac{\kappa^2}{4(e^2\vert\phi\vert^2 + 4\delta^2K^{''})}\times
\right.
\nonumber\\
&\left. \left[{\cal B} \pm\frac{\left(e^2\vert\phi\vert^2 +
4\delta^2 K^{''}\right)}{\kappa^2\,f(s)/2}
\left(e(\vert\phi\vert^2-\vert\phi_0\vert^2)-4\delta K^{\prime}
\right)\right]^2\right.\nonumber\\
&\left. + W - \frac{1}{(\kappa\,f(s))^2}\left(e^2\vert\phi\vert^2
+ 4\delta^2
K^{''}\right)\left(e(\vert\phi\vert^2-\vert\phi_0\vert^2)-4\delta
K^{\prime}
\right)^2\right.\nonumber\\
&\pm \left(e(\vert\phi\vert^2-\vert\phi_0\vert^2)-4\delta
K^{\prime} \right)\epsilon^{ij}\left[A_i +
\frac{1}{2\delta}\partial_ia\right]\partial_j{\rm log} f(s)
\nonumber\\
& \left.  \vphantom{\frac{\left(e^2\vert\phi\vert^2 + 4\delta^2
K^{''}\right)}{s/2}} \pm e\vert\phi_0\vert^2F_{xy} \right)
\label{bogomolnyi}
\end{align}
We thus see that the first three terms in eq.(\ref{bogomolnyi})
have been accommodated as perfect squares. This, together with an
appropriate choice of the potential $W$ so as to cancel the fourth
and fifth terms, would lead to a Bogomolny bound for the energy
given by the magnetic flux $\Phi$ appearing in the last term,
\begin{equation}
\Phi = \int d^2x F_{xy}
\end{equation}
or
\be \Phi = 2 \pi n \ee
for gauge fields with topological number $n\in \mathbb{Z}$. There
is however the sixth term in (\ref{bogomolnyi}) with no definite
sign preventing the obtention of a bound. Only if we put $f(s) =
1$, which corresponds to a normal Chern-Simons action for the
gauge field (see eq.(\ref{normal})), this term vanishes. In that
case one does have a bound,
\begin{equation}
{\cal E} \geq \pm e\vert\phi_0\vert^2 \Phi = 2\pi
e\vert\phi_0\vert^2  |n | \label{bound}
\end{equation}
whenever the potential is chosen as
\begin{equation}
W =
 \frac{1}{\kappa^2} \left(e^2\vert\phi\vert^2 + 4\delta^2 K^{\prime\prime} \right)
\left(e(\vert\phi\vert^2-\vert\phi_0\vert^2) - 4\delta K^{\prime}
\right)^2\label{potential}
\end{equation}
The bound is saturated by fields obeying the self-duality
equations
\begin{eqnarray}
&& D_x\phi = \mp iD_y\phi  \nonumber\\
&& D_x S = \mp iD_y S \nonumber\\
&& \kappa^2 F_{xy} = \mp 2 \left(e^2\vert\phi\vert^2 + 4\delta^2
K^{\prime\prime}  \right)
\left(e(\vert\phi\vert^2-\vert\phi_0\vert^2) - 4\delta K^\prime
\right)
\label{bpseq}
\end{eqnarray}
where the upper (lower) sign corresponds to positive (negative)
values of $\Phi$.

As it is well-known, the presence of a Chern-Simons term forces a
relation between magnetic flux and electric charge \cite{DJ}.
This makes the Chern-Simons vortices both magnetically and
electrically charged. To see this phenomenon in the present case
let us write the gauge field equation of motion, for the case
$f(s)=1$ in the form
\begin{equation}
\frac{\kappa}{2} \varepsilon^{\mu\alpha\beta}F_{\alpha\beta} =
J^\mu \label{3ese}
\end{equation}
with $J^\mu= (\rho, \vec J)$ the conserved matter current and
$\rho$ the electric charge density,
\begin{eqnarray}
  J^\mu = - 2\left(e^2 |\phi|^2 + 4K''\delta^2\right)A^\mu -
  ie(\phi\partial^\mu\phi^* + \phi^*\partial^\mu\phi) -
 4\delta
  \partial^\mu a
\end{eqnarray}
We then see that eq.(\ref{gausi}) can be rewritten in the form
\begin{equation}
\rho = -\kappa {\cal B}
\end{equation}
so that the usual relation between electric charge $\;Q = \int
\rho \; d^2x\;$  and magnetic flux in Chern-Simons theories holds,
\begin{equation}
\label{related}
Q = -\kappa \Phi
\end{equation}
Note that both the Higgs scalar and the axion contribute to the
electric charge.

\section{Supersymmetric extension}

The SUSY extension of the Chern-Simons-Higgs system with a sixth
order symmetry breaking potential was analyzed in \cite{LLW}. Let
us study now the the case in which the axion field is also
present.

Consider the following $d=3$ action, written in terms of
superfields as
\begin{eqnarray}
{\cal S}_{{\rm SUSY}} &=&-\frac{1}{2}\int d^3x\,d^2\theta \left(\frac{\kappa}{4\delta}
F(\Sigma+\Sigma^\dag){\rm Im}\left(\tilde\nabla_a\Sigma\right)W^a
\right.\hspace*{2cm}\nonumber\\
&+&\left. H(\Phi^\dag,\Phi)(\nabla^a\Phi)^\dag\nabla_a\Phi+
K_{s\bar s}(\Sigma,\Sigma^\dag)\vert \tilde\nabla_a\Sigma\vert^2 \right.\nonumber\\
&+&\left.2V(\Phi^\dag,\Phi, \Sigma,\Sigma^\dag)
\right)\label{suaction}
\end{eqnarray}
The index $a$ takes the values $1, 2$ and $\Gamma_a$, $\Phi$ and
$\Sigma$ are spinor, complex scalar and axionic superfields. Also
\begin{align}
&\nabla_a \Phi = (D_a - ie \Gamma_a)\Phi \nonumber \\
&\tilde \nabla_a \Sigma = D_a\Sigma + 2i\delta \Gamma_a \nonumber
\\
&D_a = \frac{\delta}{\delta\theta^a} + i
(\gamma^\mu)_{ab}\theta^b\partial_\mu\nonumber \end{align}
where $\mu=0,1,2$.

The lowest component of $\,\Gamma^a$, $\Phi$ and $\Sigma$ are
respectively, the gauge field, the Higgs field and $S=s + i a$
with $a$ the axion field. Concerning $F, H, K$ and $V$, they are
functionals of superfields to be fixed later. Subindices in these
functionals indicate derivatives, thus $K_{s\bar s} =
\partial_S\partial_{S^*}K=\partial_\Sigma\partial_{\bar\Sigma}
K\vert_{\theta=0}$ and so on. These functionals should be chosen
so that the supersymmetric action (\ref{suaction}) is invariant
under the supergauge transformations,
\begin{align}
\Phi \; &\to \; e^{i e \Lambda}\Phi\nonumber\\
\Gamma_a \; &\to \; \Gamma_a - D_a \Lambda \nonumber\\
\Sigma \; &\to \; \Sigma_a  - 2\delta \Lambda \label{susytrafo}
\end{align}
for any real scalar superfield $\Lambda$.
Written in the Wess-Zumino gauge, {\it i.e.},
\[\Gamma_a\vert_{\theta=0}= D^a\Gamma_a\vert_{\theta =0}= 0,\]
the spinor superfield $\Gamma_a$ is given by
\[\Gamma_a(x,\theta) =
i\theta^b(\gamma^\mu)_{ba} A_\mu(x) - 2 \theta^2 \lambda_a(x)\]
where $\lambda(x)$ is a Majorana spinor, the photino. Then, the
spinor field strength, defined as
\[W_a = \frac{1}{2} D^bD_a \Gamma_b\]
takes, in terms of component fields, the form
\[W_a(x,\theta) = \lambda_a(x) - \frac{1}{2}
\theta^b(\gamma^\mu\gamma^\nu)_{ba}F_{\mu\nu} - i
\theta^2(\gamma^\mu)^b_a \partial_\mu\lambda_b(x)\]
and satisfies the Bianchi identity,
\[D^aW_a = 0.\]
The  complex scalar superfield is defined as %
\[ \Phi(x,\theta) =
\phi(x)+ \theta^a\psi_a(x) - \theta^2 F(x) \]
where $\phi$ stands for the Higgs complex scalar field, $\psi_a$
is a Dirac bispinor, the higgsino and $F$ is a complex auxiliary
field. Finally, the superfield $\Sigma$ which contains the axion
$S$ as its lowest component, can be reduced to a complex scalar
field $\Theta$ by exponentiation,
\[\Theta = e^{\Sigma}\]
The supersymmetric transformations (\ref{susytrafo}) take
$\Gamma^a$ out from the Wess-Zumino gauge. One can however
implement a composition of SUSY and gauge transformations such
that the Wess-Zumino gauge remains valid. To do that, the new SUSY
transformation for scalar and spinorial superfields are,
respectively,
\begin{align}
\delta^{\rm WZ}_\eta \Gamma_a &= i\eta^b Q_b \Gamma_a +
D_a\tilde K\nonumber\\
\delta^{\rm WZ}_\eta \Phi &= i\eta^b Q_b \Phi +
i\tilde K\Phi\label{transf}\\
\delta^{\rm WZ}_\eta \Sigma &= i\eta^bQ_b \Sigma + i\tilde K
\nonumber
\end{align}
where $Q_a=i\partial_a+\theta^b(\gamma^\mu)_{ba}
\partial_\mu$, $\;\eta^a$ is a Majorana spinor and the real scalar
superfield $\tilde K$ is defined as,
\[\tilde K = i\theta^a(\gamma^\mu)_{ab}\eta^b A_\mu +
\theta^2\lambda^a\eta_a.\]
Let us restrict the action (\ref{suaction}) to the
case in which
\begin{equation}
F(\Sigma+\Sigma^\dag) = H(\Phi^\dag,\Phi) = 1
\end{equation}
Then, written in components, the action takes the simple form
\begin{equation}
{\cal S}_{SUSY} = {\cal S}_{\rm B} + {\cal S}_{\rm
F}
\end{equation}
with the pure bosonic action given by
\begin{equation}
{\cal S}_{\rm B}=\int d^3x\,\left\{ \frac{\kappa}
{4}\epsilon^{\mu\nu\sigma}A_\mu F_{\nu\sigma} +
\vert D_\mu\phi\vert^2 + R\vert\tilde D_\mu
S\vert^2 -\vert V_\phi\vert^2 -\frac{1}{R}\vert
V_s\vert^2\right\}
\label{boson}
\end{equation}
while  the fermionic one is given by
\begin{align}
{\cal S}_{\rm F}&=\int d^3x\,\left\{
i\bar\psi\gamma^\mu{\mathop{D}^\leftrightarrow} _\mu\psi + i
R\bar\chi\gamma^\mu {\mathop{\partial}^\leftrightarrow}_\mu\chi
%-\frac{1}{2} H_{\phi\bar
%\phi} (\bar\psi\psi)^2
-\frac{1}{2} R_{s\bar s}
(\bar\chi\chi)^2\right.\nonumber\\
&+
%\frac{1}{2}
%\left[i\bar\psi\gamma^\mu\psi(H_\phi\,D_\mu\phi -
%H_{\bar\phi} D_\mu \phi^*)\right] +
\frac{1}{2}
\left[i\bar\chi\gamma^\mu\chi(R_s\,\tilde D_\mu S
- R_{\bar s}\,\tilde D_\mu S^*)\right]\nonumber\\
&+\bar\psi\psi\left(\frac{e^2}{\kappa} \vert\phi\vert^2
%H^2
- V_{\phi\bar\phi}\right)
+\bar\chi\chi\left(\frac{4\delta^2}{\kappa}
R^2 - V_{s\bar s}\right)
%- \frac{\vert H_\phi\vert^2}{4H}
%\vert\bar\psi\psi^*\vert^2
\nonumber\\
&- \bar\psi\chi\left(\frac{2\delta e}{\kappa}\phi
%H
R + V_{\bar\phi s}\right)-
\bar\psi^*\chi^*\left(\frac{2\delta e}{\kappa}
\phi^*
%H
R + V_{\phi\bar s}\right)- \frac{\vert R_s\vert^2}
{4R}\vert\bar\chi\chi^*\vert^2\nonumber\\
&+ \frac{1}{2}\bar\psi^*\psi\left(
%\frac{H_\phi}{H}\, V_\phi
- V_{\phi\phi} -\frac{e^2}{\kappa}\phi^{*2}
%H^2
\right)+ \frac{1}{2}\bar\psi\psi^*\left(
%\frac{H_{\bar\phi}}{H}
%\,V_{\bar\phi}
- V_{\bar\phi\bar\phi}
-\frac{e^2}{\kappa}\phi^{2}
%H^2
\right)
\nonumber\\
&+ \frac{1}{2}\bar\chi^*\chi\!\!\left(\!\frac{R_s}{R}\, V_s -
V_{ss} -\frac{4\delta^2}{\kappa} R^2\!\right)\!+\! \frac{1}{2}
\bar\chi\chi^*\!\left(\!\frac{R_{\bar s}}{R} \,V_{\bar s} -
V_{\bar s\bar s} -\frac{4\delta^2}{\kappa}R^2\!\right)
\nonumber\\
&+ \left.\bar\psi^*\chi\left(\frac{2\delta e} {\kappa} \phi^*
%H
R - V_{\phi s}\right)
+\bar\psi\chi^*\left(\frac{2\delta e}{\kappa}
\phi
%H
R - V_{\bar\phi\bar s}\right) \right\},
\label{fermion}
\end{align}
where we have redefined $R\equiv K_{s\bar s}$. Here we have taken
into account that
\[
I =
\int d^3x d^2\theta\,\left( D^a\Sigma\right) W_a
\]
is a surface term which does not modify the equations of motion
\begin{align}
I &= \int d^3x d^2\theta\, D^a\left(\Sigma W_a\right) = \int
d^3x\, D^2D^a\left.\left(\Sigma
W_a\right)\right\vert_{\theta=0}\nonumber\\
&= i\int d^3x\,\partial_b^aD^b\left.\left( \Sigma
W_a\right)\right\vert_{\theta=0}\,, \nonumber
\end{align}
and can hence be neglected.

In order to extend the $N=1$ supersymmetry to an $N=2$
supersymmetry, we shall follow \cite{LLW} and allow the
transformation parameter $\eta$ to be complex, {\it i.e.},
$\eta^*\not=\eta$. Now, this is the same as making an $N=1$ SUSY
transformation followed by a $U(1)$ fermion phase rotation. Thus,
the new transformation for fermions will be $\psi_a \to
e^{i\alpha}\psi_a $ and $\psi^*_a \to e^{-i\alpha} \psi^*_a $ and
the same is valid for $\chi_a$ and $\chi^*_a$. The new SUSY
transformations act then as rotations on the fermions and one can
then see that the only terms which do not respect the extended
SUSY invariance are those on the last three lines in
(\ref{fermion}). Hence, in order to get an $N=2$ supersymmetric
model we need that,
\begin{eqnarray}
%\frac{H_\phi}{H}\, V_\phi
- V_{\phi\phi} -\frac{e^2}{\kappa}\phi^{*2}
%H^2
= 0 \; , &&\frac{2\delta e} {\kappa} \phi^*
%H
R - V_{\phi s}= 0\nonumber\\
\frac{4\delta^2}{\kappa} R^2 - V_{s\bar s }
-\frac{4\delta^2}{\kappa} R^2= 0 \;, &&\frac{2\delta e} {\kappa}
\phi
%H
R - V_{\bar\phi\bar s} = 0\nonumber
\end{eqnarray}
where $V=V(u,v)$, with $u=\phi^*\phi\;$ and $v= S + S^*$. These
equations imply that,
\[V_u = -\frac{e}{\kappa}\left(eh(u) - 2\delta r(v)\right) H (u)\]
\[ V_v = -\frac{2\delta}{\kappa}\left(2\delta r(v) - e h(u)\right)
R(v)\]
where
\begin{equation}
\frac{d}{dv}r(v) = R(v) \; , \;\;\; \frac{d}
{du}h(u) = 1.
\end{equation}
We obviously have $h=u-u_0 = \vert\phi\vert^2
-\vert\phi_0\vert^2$, and since $S+S^*=2s$, then $R =
K^{\prime\prime}$ and $r = 2 K^\prime$ where primes stand for
derivatives with respect to $s$. From (\ref{boson}) the potential
is,
\[W =
%\frac{1}{H}
\vert V_\phi\vert^2 + \frac{1} {R}\vert V_s\vert^2 =
\frac{1}{\kappa^2} (e^2\vert \phi\vert^2 +
4\delta^2K^{\prime\prime}) \left[e(\vert\phi\vert^2 -
\vert\phi_0\vert^2) - 4\delta K^\prime\right]^2\]
which is exactly what we obtained in (\ref{potential}).\par
\noindent In order to get the Bogomolny bound and the self-dual
equations one can analyze  the supercharge algebra as in
\cite{ens1}. Alternatively, one can directly consider the
component field SUSY transformations    $(\delta_\eta X =
\eta^a\delta_a X$),
\begin{align}
\delta_a A_\mu&=-i (\gamma_\mu)_a^b \lambda_b \,,
&\delta_a\lambda_b &= \frac{1}{2}
(\gamma^\mu\gamma^\nu)_{ab}F_{\mu\nu} \nonumber\\
\delta_a\phi&= - \psi_a \,, &\delta_a S &=- \chi_a \nonumber\\
\delta_a\psi_b&= \epsilon_{ab} F - i (\gamma^\mu)_{ab} D_\mu\phi
\,,&
\delta_a\chi_b &=\epsilon_{ab} J - i(\gamma^\mu)_{ab}\tilde D_\mu S \nonumber\\
\delta_a F&= i (\gamma^\mu)_a^b D_\mu\psi_b + 2\lambda_a\phi \,, &
\delta_a J &=i (\gamma^\mu)_a^b\partial_\mu\chi_b + 2\lambda_a
\label{doblecol}
\end{align}
and their complex conjugated ($\delta X^\dag = \eta^{*a}\delta_a
X^\dag$) and reobtain Bogomolny equations  (\ref{bpseq}) just by
putting all fermion fields to zero in (\ref{doblecol}) and then
ask the SUSY transformations for $\psi_a$ and $\chi_a$ to vanish
once the auxiliary fields have been written in terms of dynamical
fields using their equations of motion. The first condition
corresponds to a restriction to the bosonic sector, the second one
implies that physical states are supersymmetry invariant.

%%%%%%%%%%%%%%%%%%%%%%%%%%%%%%%%%%%%%%%%%%%%%%%%%%%%%%%%%%%%%%%%%%%%%%%%%
\section{Vortex-like solutions}

We present in this section some vortex solutions to the BPS
equations of motion. We choose for the Kh\"aler potential the form
\begin{equation} K=-M^2\log\left(S+S^*\right) \end{equation}

As in \cite{Blanco}, we shall analyze separately two cases: first,
we consider the case in which the vortex is supported by the Higgs
field (``$\phi$-strings'' solutions) , in the sense that at
infinity it behaves as in the ordinary Nielsen-Olesen vortex, with
its winding number linked to the magnetic flux. Then, we shall
consider the case in which the vortex is supported by the $s$
field, a solution that we shall call  an ``s-string''. In this
case it is the axion winding number which is related to the
magnetic flux.

In the first case, in order to obtain  $\phi$-string solutions we
make the ansatz \cite{Blanco}
\begin{equation}
\phi_1=f(r)e^{in\theta} \ \ \ \ \ \ \ \ \ S=s(r)-2i\delta m\theta
\ \ \ \ \ \ \ \ \ A_{\theta}=n\frac{v(r)}{r} \label{ansatz}
\end{equation}
where $n$ is the topological charge of the Higgs and $m$ is the
topological charge of the axion.

It is convenient to work with dimensionless variables by defining
\begin{align}
&\tau=\alpha r,\ \ \ \ \ \ \ \alpha=\frac{e^2\phi_0^2}{\kappa} \nonumber\\
&x(\tau)=v(\tau/\alpha) \nonumber\\
&y(\tau)=e\delta^{-1} s(\tau/\alpha) \nonumber\\
&z(\tau)=\phi_0^{-1} f(\tau/\alpha)
\end{align}
With this convention the equations read,
\begin{align}
x'&=-\frac{2\tau}{\left|n\right|} \left(z^2 +
\frac{4\beta}{y^2}\right) \left(z^2 - 1 + \frac{4\beta}{y}\right)
\label{eqx}\\
y'&=-\frac{2}{\tau}(\left|m\right|-\left|n\right|x) \label{eqy}\\
z'&=\frac{z}{\tau}(1-x)\left|n\right| \label{eqz}
\end{align}
where $\beta=M^2/\phi_0^2$. From the first two equations we can
integrate $y(\tau)$ in terms of $z(\tau)$, obtaining
\begin{equation}
y(\tau)=2(\left|n\right|-\left|m\right|)\log\tau-2\log z(\tau)+k
\label{soly}
\end{equation}
with $k$ an arbitrary integration constant. Thus, we end with a
system of two first-order coupled differential equations for
$x(\tau)$ and $z(\tau)$.

The boundary conditions for equations \eqref{eqx} and \eqref{eqz}
can be determined as follows. By equation \eqref{ansatz}, the
function $x(\tau)$ must vanish at the origin, so eq. \eqref{eqz}
implies that $z(\tau)$ also vanishes at the origin as
$\tau^{|n|}$. Indeed, eqs. \eqref{eqx} and \eqref{soly} imply that
$x\sim \tau^2$ for $m=0$ and $x\sim \tau^2/\log(\tau)^2$.

For large $\tau$, the function $x$ tends to $1$, thus $z$ also
tends to $1$ unless $|n|=|m|$. In this last case, $z\, \to \,
z_0$, where $z_0$ is the solution of the algebraic equation
\be z_0^2 + \frac{4 \beta}{k-2 \log z_0 } =1 \ee

\begin{figure}[h]
\begin{center}
\includegraphics[angle=0, width=10cm]{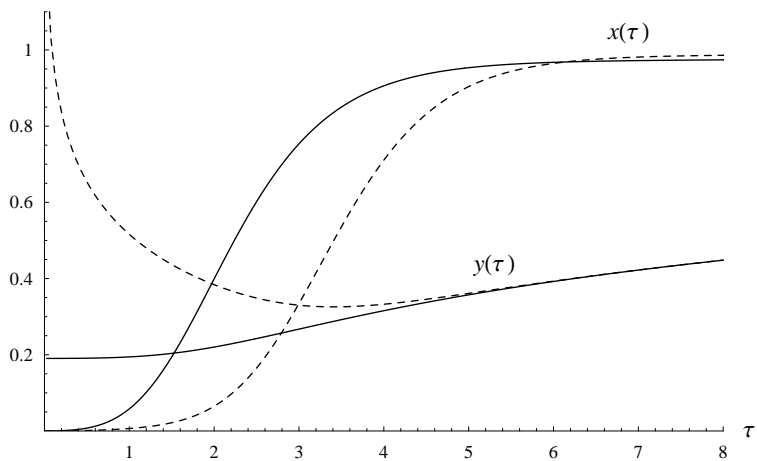}
\end{center}
\smallskip
\caption{ Plot of the gauge field $x(\tau)$ and the dilaton
$y(\tau)$ for $n=1$ and $m=0$ (solid line) and for $n=2$ and $m=1$
(dashed line). In both cases $\beta=1$ and $k=10$. }
\label{fig-xy}
\end{figure}

To solve the differential equations we employ a relaxation method
for boundary values problem. Such a method determines the solution
by starting with an initial guess and improving it iteratively. We
present in figure \ref{fig-xy} some solution profiles for the
gauge field $x(\tau)$ and for the dilaton $y(\tau)$.

%%%%%%%%%%%%%%%%%%%%%%%%%%%%%%%%%%%%%%%%%%%%%%%%%%%%%%%%%%%%%%%%%%%%%%%%%%%%
%%%%%%%%%%%%%%%%%%%%%%%%%%%%%%%%%%%%%%%%%%%%%%%%%%%%%%%%%%%%%%%%%%%%%%%%%%%%%

\begin{figure}[h]
\begin{center}
\includegraphics[angle=0, width=10cm]{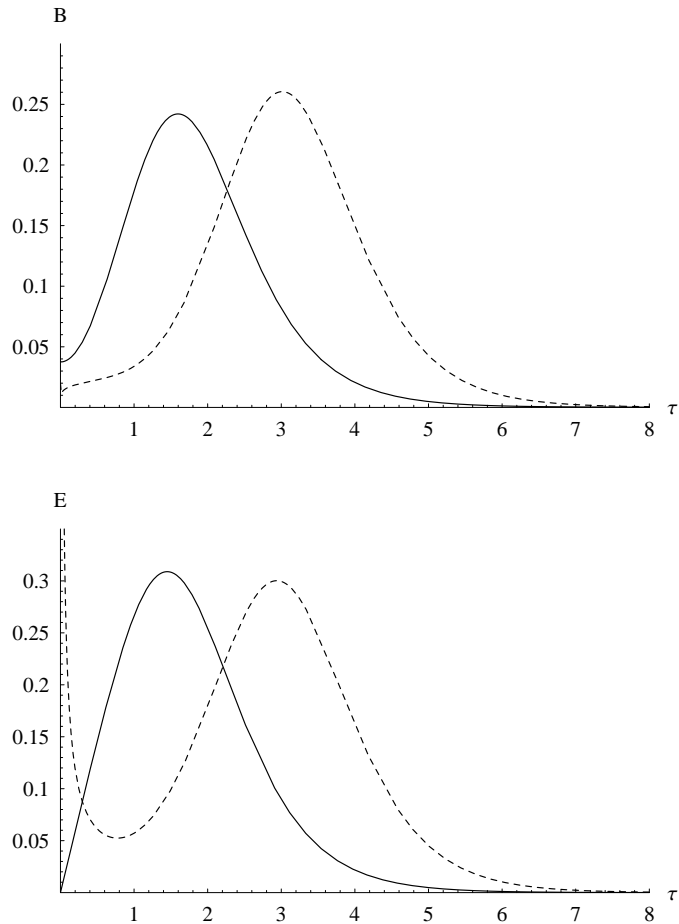}
\end{center}
\smallskip
\caption{ Vortex magnetic and electric fields. The solid line
corresponds to $n=1$ and $m=0$ and the dashed line to $n=2,\;
m=1$. In both cases $\beta=1$ and $k=10$.}
\label{magnetic-electric}
\end{figure}

Already from ansatz (\ref{ansatz}) one can see that the magnetic
flux and electric charge ar quantized according to
\begin{equation}
\Phi = \frac{2\pi}{e} n   \; , \;\;\; Q =-\frac{2\pi\kappa}{e} n
\end{equation}
There is an interesting property, typical of Chern-Simons vortices
that also holds in our model: both the magnetic field and the
(radial) electric field are concentrated in rings surrounding the
zeroes of the Higgs field. We show this behavior in figure
\ref{magnetic-electric}.

%%%%%%%%%%%%%%%%%%%%%%%%%%%%%%%%%%%%%%%%%%%%%%%%%%%%%%%%%%%%%%%%%%%%%%%%%%%%%
%%%%%%%%%%%%%%%%%%%%%%%%%%%%%%%%%%%%%%%%%%%%%%%%%%%%%%%%%%%%%%%%%%%%%%%%%%%%%

\begin{figure}[h]
\begin{center}
\includegraphics[angle=0, width=10cm]{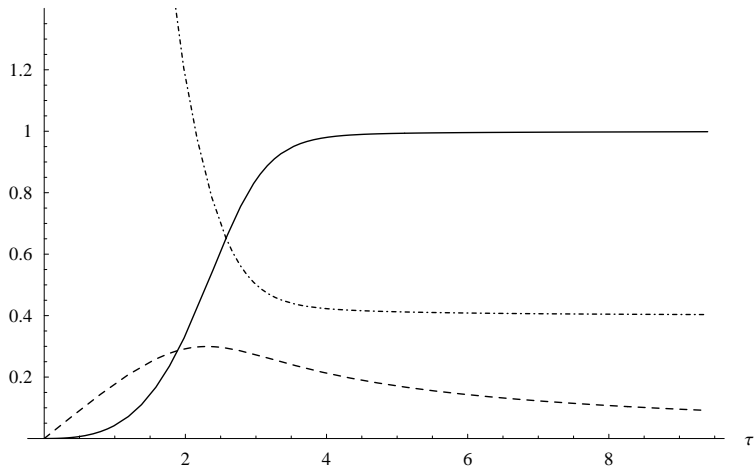}
\end{center}
\smallskip
\caption{ Profile of an s-string solution. The solid line
corresponds to the gauge field $x(\tau)$, the dot-dash line to the
dilaton $y(\tau)$ and the dashed line to the Higgs field
$z(\tau)$. For this solution $n=1, \; m=2$, $\beta=0.1$, and
$k=0.1$. } \label{fig-x-s-string}
\end{figure}

We also solved the BPS equations when the field $s$ tends
asymptotically to a constants (``s-string", \cite{Blanco}). In
this case the magnetic charge is equal to the topological charge
of the axion field, so we have as an appropriate ansatz,
\begin{equation}
\phi=f(r)e^{in\theta} \ \ \ \ \ \ \ \ \ S=s(r)-2i\delta m\theta
\ \ \ \ \ \ \ \ \ A_{\theta}=m\frac{v(r)}{r}
\end{equation}
With this ansatz the BPS equations take the form
\begin{align}
x'&=-\frac{2\tau}{\left|m\right|} \left(z^2 +
\frac{4\beta}{y^2}\right) \left(z^2 - 1 + \frac{4\beta}{y}\right)
\label{eqsx}\\
y'&=-\frac{2}{\tau}\left|m\right|\left(1-x \right) \label{eqsy}\\
z'&=\frac{z}{\tau}\left(\left|n\right| - \left|m\right| x\right)
\label{eqsz}
\end{align}
Again, we can integrate $y(\tau)$ in terms of $z(\tau)$, obtaining
\begin{equation}
y(\tau)=2(\left|n\right|-\left|m\right|)\log\tau-2\log z(\tau)+k
\end{equation}
We see from this equation that for consistency, $z(\tau) \sim
\tau^{n-m}$ for $\tau \to \infty$ in contrast with what happens
for the $\phi$-string. Concerning the gauge field boundary
condition, one has $\lim_{\tau \to \infty}x(\tau) = 1$. We present
in figure \ref{fig-x-s-string} some profile of the resulting
s-string solutions.

%%%%%%%%%%%%%%%%%%%%%%%%%%%%%%%%%%%%%%%%%%%%%%%%%%%%%%%%%%%%%%%%%%%%%%%%%%%%
%%%%%%%%%%%%%%%%%%%%%%%%%%%%%%%%%%%%%%%%%%%%%%%%%%%%%%%%%%%%%%%%%%%%%%%%%%%%%

\section{Discussion}
We have been able to write first order Bogomolny equations for the
Chern-Simons-Higgs-axion system provided the gauge field dynamics
is governed by a normal Chern-Simons action and the potential
takes the form (\ref{potential}), with coupling constants
satisfying the usual relation. As in the $3+1$ case discussed in
\cite{Blanco}, the first order equation for the scalar $S$
containing the axion field has the same form as the ordinary
Bogomolny equation for the Higgs field. The sixth order symmetry
breaking potential (necessary to accommodate the energy as a sum
of perfect squares plus a topological term) is modified due to the
presence of the axion. As it always happens when matter is coupled
to a Chern-Simons term, magnetic flux and electric charge turn to
be proportional (see eq.(\ref{related})). Were we considering
non-Abelian gauge fields, then the proportionality constant
$\kappa$ should be quantized, implying in turn quantization of
both electric charge and magnetic flux at the classical level.

The same constraints and results  were found by constructing the
SUSY extension of the purely bosonic model. In this case we have
seen that the conditions on the potential and coupling constants
arise when extending supersymmetry from $N=1$ to $N=2$ while the
BPS equations can be inferred by asking physical states to be
annihilated by SUSY generators.

The numerical solutions to the BPS equations presented in section
4 show that the axionless string solutions found in
\cite{Hong}-\cite{JW} are not much modified by the axion which,
however, contributes to the electric charge of the string
configuration in a clear way. These explicit solutions could be of
relevance in the context of cosmic strings and, due to the
coupling to the axion and their electric charge, their dynamics
could be very different from that usually considered. This, and
the extension to the non-Abelian case (where both the magnetic
flux and the electric charge are quantized) are issues that we
hope to address in a future work.

\section{Appendix: Conventions}
We use the following representation of the $\gamma$ matrices,
\[ (\gamma^0)_a^b = (\sigma^2)_a^b,\quad (\gamma^1)_a^b=
(i\sigma^3)_a^b,\quad (\gamma^2)_a^b=(i\sigma^1)_a^b
\]
Thus,
\[\gamma^\mu\gamma^\nu = g^{\mu\nu} -
i\epsilon^{\mu\nu\sigma} \gamma_\sigma\]
The ``metric'' in spinor space is given by
\begin{equation}
C_{ab} = i(\sigma^2)_{ab} = \left(
\begin{array}{c c}
0  & 1\\
-1 & 0
\end{array}
\right) = C^{ab}
\end{equation}
So,
\[C^{ab}C_{cd} = \delta^a_{[c}\delta^b_{d]}\]
In particular,
\be \psi^a = C^{ab}\psi_b \,, \quad\quad \psi_a= \psi^b C_{ba} \,,
\quad\quad \theta^2 = \frac{1}{2}\theta^a\theta_a =
\theta_2\theta_1 \ee
We define the integration measure in the superspace,
\[ \int
d^2\theta \; \theta^2 = -1 \]
Also, useful formulae are,
\[\theta_a\theta_b = -C_{ab}
\theta^2,\quad\quad\theta^a\theta^b=-C^{ac}\theta^2,\quad
\theta^a\theta_b = \delta^a_b\theta^2= -\theta_b\theta^a\]
\[A^{[a}B^{b]} = -C^{ab} A^cB_c,\quad A_{[a}B_{b]} = -C_{ab}
A^cB_c\]
The $\gamma$ matrices are real and symmetric when lowering their
indices, {\it i.e.},
\[(\gamma^0)_{ab} = (\gamma^0)_a^c C_{cb} =i I_{ab} (= i
\delta_{ab}),\quad (\gamma^1)_{ab} = (\gamma^1)_a^cC_{cb} =
i(\sigma^1)_{ab},\]
\[(\gamma^2)_{ab} = (\gamma^2)_a^cC_{cb} = -i(\sigma^3)_{ab}\]
Then, any vectorial representation can be written in terms of the
$\gamma$ matrices with spinorial indices, {\it i.e.},
\[V_{ab} = (\gamma^\mu)_{ab} V_\mu\]
where, $\gamma_{ab}$ are imaginary and symmetric $2\times 2$
matrices. Also, we can define the spacetime derivative in terms of
spinorial indices,
\[\partial_{ab} = (\gamma^\mu)_{ab}\partial_\mu\]
The spinorial derivative is define as,
\begin{equation}
\partial_a = \frac{\partial}{\partial\theta^a},\quad \partial^a=
-\frac{\partial}{\partial\theta_a}\label{def1}\end{equation}
\begin{equation}
\partial_{ab}= \frac{\partial}{\partial x^{ab}},\quad\partial^a_b
= -\frac{\partial}{\partial x^b_a}\label{def2}\end{equation}
Then, the supersymmetric covariant derivative can be written in
terms of this notation as,
\[D_a = \partial_a + i\theta^b\partial_{ba}\]
Note that the derivatives $\frac{\partial}{\partial \theta^a}$ and
$\frac{\partial}{\partial x^{ab}}$ do not raise and lower indices
in the same way as spinors,
\[ \partial_a = (\partial_a\theta_b)\frac{\partial}
{\partial\theta_b} = C_{db}
(\partial_a\theta^d)\frac{\partial}{\partial\theta_b} =
C_{ab}\frac{\partial}{\partial\theta_b}=C_{ba}
\partial^b\]
\[\partial_{ba}= \frac{\partial x^c_d}{\partial x^{ba}}
\frac{\partial}{\partial x^c_d} = C_{ed} \frac{\partial x^{ce}}
{\partial x^{ba}}\frac{\partial}{\partial x^c_d} = C_{ad}
\frac{\partial}{\partial x^b_d} = C_{da}\partial_b^d \]
causing the appearance of a minus sign in the definitions of
$\partial^a$ and $\partial^a_b$ in (\ref{def1}) and (\ref{def2}).

%%%%%%%%%%%%%%%%%%%%%%%%%%%%%%%%%%%%%%%%%%%%%%%%%%%%%%%%%%%%%%%%%%%%%%%%%%%%%
%%%%%%%%%%%%%%%%%%%%%%%%%%%%%%%%%%%%%%%%%%%%%%%%%%%%%%%%%%%%%%%%%%%%%%%%%%%%%
%%%%%%%%%%%%%%%%%%%%%%%%%%%%%%%%%%%%%%%%%%%%%%%%%%%%%%%%%%%%%%%%%%%%%%%%%%%%%

\begin{acknowledgments}
The work of J.L-S was supported in part by Proyecto FONDECYT 3060
002. The work of F.A.S. was supported in part by grants 6160
PIP-CONICET, CICBA and X319 UNLP. J.L-S would like to thank La
Plata University and USACH, for hospitality while this work was
been performed.
\end{acknowledgments}

%%%%%%%%%%%%%%%%%%%%%%%%%%%%%%%%%%%%%%%%%%%%%%%%%%%%%%%%%%%%%%%%%%%%%%%%%%%%%
%%%%%%%%%%%%%%%%%%%%%%%%%%%%%%%%%%%%%%%%%%%%%%%%%%%%%%%%%%%%%%%%%%%%%%%%%%%%%
%%%%%%%%%%%%%%%%%%%%%%%%%%%%%%%%%%%%%%%%%%%%%%%%%%%%%%%%%%%%%%%%%%%%%%%%%%%%%

%%%%%%%%%%%%%%%%%%%%%%%%%%%%%%%%%%%%%%%%%%%%%%%%%%%%%%%%%%%%%%%%%%%%%%%%%%%%%
%%%%%%%%%%%%%%%%%%%%%%%%%%%%%%%%%%%%%%%%%%%%%%%%%%%%%%%%%%%%%%%%%%%%%%%%%%%%%


\begin{thebibliography}{99}
\bibitem{witt} E.~Witten,
  %``Cosmic Superstrings,''
  Phys.\ Lett.\ B {\bf 153} (1985) 243.
\bibitem{pol}
  J.~Polchinski,
  %``Introduction to cosmic F- and D-strings,''
  arXiv:hep-th/0412244.
\bibitem{ens1} J.~D.~Edelstein, C.~N\'u\~nez and F.~Schaposnik,
  %``Supersymmetry and Bogomolny equations in the Abelian Higgs model,''
  Phys.\ Lett.\ B {\bf 329} (1994) 39.
\bibitem{ens2}
  J.~D.~Edelstein, C.~N\'u\~nez and F.~A.~Schaposnik,
  %``Supergravity and a Bogomolny bound in three-dimensions,''
  Nucl.\ Phys.\ B {\bf 458} (1996) 165.
\bibitem{ens3}
  J.~D.~Edelstein, C.~N\'u\~nez and F.~A.~Schaposnik,
  %``Bogomol'nyi Bounds and Killing Spinors in d=3 Supergravity,''
  Phys.\ Lett.\ B {\bf 375} (1996) 163.
\bibitem{Shifman:2003uh}
  M.~Shifman and A.~Yung,
  %``Localization of non-Abelian gauge fields on domain walls at weak  coupling
  %(D-brane prototypes II),''
  Phys.\ Rev.\ D {\bf 70} (2004) 025013.
\bibitem{Dvali1}
  G.~Dvali, R.~Kallosh and A.~Van Proeyen,
  %``D-term strings,''
  JHEP {\bf 0401} (2004) 035.
\bibitem{Binetruy}
  P.~Binetruy, G.~Dvali, R.~Kallosh and A.~Van Proeyen,
  %``Fayet-Iliopoulos terms in supergravity and cosmology,''
  Class.\ Quant.\ Grav.\  {\bf 21} (2004) 3137.
\bibitem{Achucarro:2004ry}
  A.~Achucarro and J.~Urrestilla,
  %``F-term strings in the Bogomolnyi limit are also BPS states,''
  JHEP {\bf 0408} (2004) 050.
\bibitem{Jeannerot:2004bt}
  R.~Jeannerot and M.~Postma,
  %``Chiral cosmic strings in supergravity,''
  JHEP {\bf 0412} (2004) 043.
\bibitem{Gubser:2004tf}
  S.~S.~Gubser, C.~P.~Herzog and I.~R.~Klebanov,
  %``Variations on the warped deformed conifold,''
  Comptes Rendus Physique {\bf 5} (2004) 1031.
\bibitem{Blanco}J.~J.~Blanco-Pillado, G.~Dvali and M.~Redi,
  %``Cosmic D-strings as axionic D-term strings,''
  Phys.\ Rev.\ D {\bf 72} (2005) 105002.
\bibitem{Parameswaran:2005mm}
  S.~L.~Parameswaran, G.~Tasinato and I.~Zavala,
  %``The 6D SuperSwirl,''
  arXiv:hep-th/0509061.
\bibitem{Avz}
  A.~Achucarro, A.~Celi, M.~Esole, J.~Van den Bergh and A.~Van Proeyen,
  %``D-term cosmic strings from N = 2 supergravity,''
  arXiv:hep-th/0511001.
\bibitem{Auzzi}
  R.~Auzzi, M.~Shifman and A.~Yung,
  %``Composite non-Abelian flux tubes in N = 2 SQCD,''
  arXiv:hep-th/0511150.
\bibitem{Gorsky}
  A.~Gorsky, M.~Shifman and A.~Yung,
  %``Nonabelian strings and axion,''
  arXiv:hep-th/0601131.
  \bibitem{Hong}
  J.~Hong, Y.~Kim and P.~Y.~Pac,
  %``On The Multivortex Solutions Of The Abelian Chern-Simons-Higgs Theory,''
  Phys.\ Rev.\ Lett.\  {\bf 64} (1990) 2230.
  \bibitem{JW}
  R.~Jackiw and E.~J.~Weinberg,
  %``Selfdual Chern-Simons Vortices,''
  Phys.\ Rev.\ Lett.\  {\bf 64} (1990) 2234.
  \bibitem{JZ}
  B.~Julia and A.~Zee,
  %``Poles With Both Magnetic And Electric Charges In Nonabelian Gauge Theory,''
  Phys.\ Rev.\ D {\bf 11}, 2227 (1975).
  \bibitem{Burgess}
  M.~Burgess,
  %``Chern-Simons vortices in an open system,''
  Phys.\ Rev.\ D {\bf 52} (1995) 1165
\bibitem{FG}
  W.~Garcia Fuertes and J.~Mateos Guilarte,
  %``Self-dual solitons in N = 2 supersymmetric Chern-Simons gauge theory,''
  J.\ Math.\ Phys.\  {\bf 38} (1997) 6214.
\bibitem{DJ}
  S.~Deser, R.~Jackiw and S.~Templeton,
  %``Three-Dimensional Massive Gauge Theories,''
  Phys.\ Rev.\ Lett.\  {\bf 48} (1982) 975;
 S.~Deser, R.~Jackiw and S.~Templeton,
  %``Topologically Massive Gauge Theories,''
  Annals Phys.\  {\bf 140}, 372 (1982)
\bibitem{LLW}
  C.~k.~Lee, K.~M.~Lee and E.~J.~Weinberg,
  %``Supersymmetry And Selfdual Chern-Simons Systems,''
  Phys.\ Lett.\ B {\bf 243} (1990) 105.
\end{thebibliography}
\end{document}